\pdfoutput=1 

\newif\ifreview
\reviewfalse

\ifreview
    \documentclass[a4paper,onecolumn]{article}
    \usepackage{lineno, setspace}
\else
    \documentclass[a4paper,twocolumn]{article}
\fi

\usepackage{IEK10} 
\usepackage[numbers,square]{natbib}

\usepackage{threeparttable}

\usepackage{upgreek}
\usepackage{bbm}

\DeclareSIUnit\year{yr}
\DeclareSIUnit\annum{a}
\DeclareSIUnit\ton{t}
\DeclareSIUnit{\million}{\text{M}}

\def\IEK10{
  Institute of Climate and Energy Systems,
  Energy Systems Engineering (ICE-1),
  Forschungszentrum J\"ulich GmbH,
  J\"ulich 52425,
  Germany
}
\def\RWTH{
  RWTH Aachen University,
  Aachen 52062,
  Germany
}
\def\JARAENERGY{
  JARA-ENERGY,
  J{\"u}lich 52425,
  Germany
}
\def\JARACSD{
  JARA-CSD,
  J{\"u}lich 52425,
  Germany
}
\def\SVT{
  RWTH Aachen University,
  Process Systems Engineering (AVT.SVT),
  Aachen 52074,
  Germany
}

\newcommand{\mytitle}{CHAMB-GA: A Containerized HPC Scalable Microservice-Based Framework for Genetic Algorithms}

\newcommand{\affil}{
  \begin{itemize}[leftmargin=3mm, itemsep=0mm]
    \item[$^a$]\IEK10
    \item[$^b$]\RWTH
    \item[$^c$]\SVT
    \item[$^d$]\JARAENERGY
    \item[$^e$]\JARACSD
  \end{itemize}
}

\def\firstAuthor{Felix Bonhoff}
\newcommand{\myauthor}{\firstAuthor$^{a,b}$,
Thiemo Pesch$^{a}$,
Andrea Benigni$^{d,a}$,
Alexander Mitsos$^{e,a,c}$,
Manuel Dahmen$^{a,*}$
}
\newcommand{\correspondingauthor}{M. Dahmen, \IEK10 \newline E-mail: m.dahmen@fz-juelich.de}
\author{\myauthor}

\usepackage[
  colorlinks,
  linkcolor=blue,
  citecolor=blue,
  urlcolor=blue,
  pdftitle={\mytitle},
  pdfauthor={\firstAuthor}
]{hyperref}
\usepackage[capitalise, nameinlink]{cleveref}
\crefname{table}{Tab.}{Tab.}

\usepackage[colorinlistoftodos]{todonotes}

\usepackage{cuted}

\begin{document}
\ifreview\else
\twocolumn[
\begin{@twocolumnfalse}
\fi

  \thispagestyle{firststyle}

  \begin{center}
    \begin{large}
      \textbf{\mytitle}
    \end{large} \\
    \myauthor
  \end{center}

  \vspace{0.5cm}
  \begin{footnotesize}
    \affil
  \end{footnotesize}
  \vspace{0.5cm}
Metaheuristic-based global optimization with embedded, long-running simulations is a computationally expensive process. 
To support various stages of development and execution, a seamless transition from personal computers to distributed clusters is desired, enabling execution across all computational scales. However, existing tool chains are often characterized by rigidity and hardware-bound constraints, which impede scalability and the integration of complex simulations.
Bridging this gap, we present a \underline{c}ontainerized \underline{H}PC sc\underline{a}lable \underline{m}icroservice-\underline{b}ased framework for \underline{g}enetic \underline{a}lgorithms with embedded simulations (CHAMB-GA). The deployment of the framework scales consistently across cloud infrastructure via container orchestration and HPC clusters via batch-scheduled parallel execution. Users provide the GA operators and simulation backend separately. The framework is designed to run these components in a distributed and decoupled manner, mapped to separate hardware. This approach ensures that the fitness evaluation and genetic operations are not managed within the same process and are utilizing distinct parts of the compute infrastructure. 
A central message broker coordinates asynchronous manager-worker communication between microservices, thereby parallelizing evolutionary operations and fitness evaluations. We demonstrate CHAMB-GA's scalability, portability, and reproducibility, while facilitating the integration of external tools and complex simulations on benchmark and powerflow problems.
The capabilities of CHAMB-GA are validated in a two-part approach: (i) a benchmark study demonstrating minimal overhead while scaling to over 3,500 CPU cores, and (ii) a dispatch optimization of High Voltage Direct Current (HVDC) lines in the German transmission grid, showing seamless migration from Kubernetes to SLURM, combined horizontal and vertical scaling, and integration of multi-stage workflows.
\vspace*{5mm}


\ifreview\else
\end{@twocolumnfalse}
]
\fi

\section{Introduction}
Finding (near) optimal solutions to many real optimization problems is computationally challenging and may include embedded simulations, by choice or necessity. For instance, in the power systems domain, a prototypical simulation task is solving the AC powerflow equations. The computational effort for such simulations rapidly increases as network size expands~\citep{shinHierarchicalOptimizationArchitecture2019, bottcherSolvingACPower2023}, which becomes especially critical when many simulations have to be performed, for example in the context of security assessment, contingency analysis, or probabilistic studies. 
Furthermore, future approaches that integrate additional energy sectors, such as heating and gas~\citep{dibosHeatNetSimOpensourceSimulation2024,luGasNetSimOpenSourcePackage2022}, are expected to further increase model complexity, thereby making the associated simulation problems more computationally demanding.

Solving optimization problems with embedded simulations is often approached with heuristic solvers~\citep{chughSurveyHandlingComputationally2019,kaushikOptimalPlacementRenewable2022}, which require extensive sampling of the search space to limit the risk of being trapped in a suboptimal solution~\citep{mallipeddiEmpiricalStudyEffect2008}.

Many population-based heuristic algorithms, like genetic algorithms (GAs), inherently involve numerous independent fitness evaluations, making them specifically well-suited to run on parallel hardware.
Inspired by natural selection, GAs are a class of meta heuristic optimization schemes that iteratively evolve a population of candidate solutions, called individuals, toward an optimal state. This process is driven by a few core mechanisms: evaluating the solution quality of each individual (fitness) via a problem-specific fitness function, selecting the most promising individuals (selection), and applying genetic operations (crossover and mutation) to generate the population of the subsequent iteration (generation)~\citep{talbiMetaheuristicsDesignImplementation2009}. 

The growing accessibility and performance of parallel computing infrastructure, such as multicore systems, GPUs, and distributed architectures like HPC systems or cloud infrastructures, align with the increasing computational demands of large populations and complexity of fitness evaluations in GAs~\citep{dongarraHardwareTrendsImpacting2024,vargheseNextGenerationCloud2018}. In combination with being applicable to a wide variety of problems, GAs provide a flexible environment to utilize computational power and quickly adapt to different fitness evaluations and genetic operations. 
Particularly in research contexts, the ease of use, modularity, compatibility with complex simulation tools used for fitness evaluation, and integration with machine learning workflows to aid evolution~\citep{maAutomatedAlgorithmDesign2026}, make GAs a popular choice for solving a wide range of optimization problems~\citep{chughSurveyHandlingComputationally2019,kaushikOptimalPlacementRenewable2022}.

Many approaches and solutions arose to utilize this inherent parallelism, often closely coupled to current developments in computing hardware, such as multi-core processors, GPUs, cloud-computing and HPC systems.~\citep{ivanovicEfficientEvolutionaryOptimization2022,gongDistributedEvolutionaryAlgorithms2015}

GAs are a widely used class of evolutionary algorithms (EAs), a generalization combining all stochastic search methods that mimic natural evolution. ~\citep{cantu-pazSurveyParallelGenetic1998} defined a taxonomy of three types of parallelism in EAs: (i) The global model (manager-worker model) consisting of a single global population distributed across multiple worker processes. (ii) The coarse-grained model (island model) which partitions the population into multiple subpopulations, or islands, that evolve mostly independently of each other.
Occasional migration of few individuals between islands to exchange knowledge enhances exploration. (iii) The fine-grained model where individuals undergo an evolutionary process on a structured grid, i.e., each individual only interacts with its neighbours or individuals from a neighbouring region, depending on the algorithm settings.

To combine large-scale scalability with modular integration of different software stacks and simulation tools across container orchestration software and batch scheduling alike, we present a \underline{c}ontainerized \underline{H}PC sc\underline{a}lable \underline{m}icroservice \underline{b}ased framework for \underline{g}enetic \underline{a}lgorithms with embedded simulations (CHAMB-GA). CHAMB-GA is a lightweight, open-source framework specifically designed to support scalable execution of evolutionary algorithms driven by computationally expensive fitness evaluations in various distributed compute environments. Utilizing containerized components enables independent development of separate tasks, such as propagation and fitness evaluation, and portability between different hardware, promoting reproducibility~\citep{brinckmanComputingEnvironmentsReproducibility2019,zhangSingularitybasedContainerTechnology2017}. Specifically, CHAMB-GA is a set of orchestration scripts that at its core employs a RabbitMQ~\citep{RabbitMQOneBroker2025,RabbitmqRabbitmqserver2026} message broker. Rather than adding more broker nodes, RabbitMQ scales vertically with the available CPU cores~\citep{RabbitMQOneBroker2025,stenmanBeamBookUnderstanding2025} achieving good scalability while being deployable independently of the chosen orchestration software. The advanced routing capabilities of the broker~\citep{RabbitMQOneBroker2025} are used to internally handle parallel schemes such as the manager-worker scheme, the island model, or multi-stage workflows, as well as dynamically adjust worker counts and handling load balancing without redeployment or relying on a fixed pattern. 
All applications and tools run in containerized environments and can be combined to interact with each other in a scalable manner, enabling, e.g.,~multi-oracle analysis, see, e.g.,~\citet{yaochujinFrameworkEvolutionaryOptimization2002}, or hyperparameter tuning.
Various works have combined GA inherent parallelization methods with distributed and parallel hardware~\citep{khalloofGenericFlexibleScalable2023,biscaniParallelGlobalMultiobjective2020,salzaPasqualesalzaElephant562021}. We qualitatively compare CHAMB-GA to these approaches in Section 2.

The remainder of this paper is structured as follows. Section 2 provides an overview of related works. Section 3 introduces the CHAMB-GA framework and describes key design choices and features. In Section 4, scalability and overhead are analyzed, followed by a power systems optimization problem demonstrating flexible resource assignment and a multi-stage workflow. Section 5 concludes with final remarks and possible future work.

\begin{table*}[ht]
    \centering
    \caption{Comparison of evolutionary computing frameworks.}
    \label{tab:framework_comparison}
    \begin{tabularx}{\textwidth}{
        >{\raggedright\arraybackslash}p{2cm}
        >{\raggedright\arraybackslash}p{3.1cm}
        >{\raggedright\arraybackslash}X
        >{\raggedright\arraybackslash}X
    }
        \toprule
        \textbf{Class} & \textbf{Relevant frameworks} & \textbf{Main features } & \textbf{Existing frameworks distinctions to CHAMB-GA} \\
        \midrule
        Monolithic &
        \begin{itemize}[left=0pt,label={\tiny\textbullet},itemsep=0pt,topsep=0pt,partopsep=0pt,nosep]
            \item DEAP~\citep{derainvilleDEAPPythonFramework2012}
            \item Geneva~\citep{berlichGenevaManual2025}
            \item Pagmo~\citep{biscaniParallelGlobalMultiobjective2020}
        \end{itemize} &
        Highly integrated using lower-level parallelism (MPI, multiprocessing, and streaming multiprocessors) &
        \begin{itemize}[left=0pt,label={\tiny\textbullet},itemsep=0pt,topsep=0pt,partopsep=0pt,nosep]
            \item Domain and language specific
            \item No flexible integration of external tools
            \item No native support for heterogeneous hardware
        \end{itemize} \\
        \addlinespace
        Functional / data-flow &
        \begin{itemize}[left=0pt,label={\tiny\textbullet},itemsep=0pt,topsep=0pt,partopsep=0pt,nosep]
            \item FlexGP~\citep{sherryFlexGPGeneticProgramming2012}
            \item Offspring~\citep{vecchiolaMultiObjectiveProblemSolving2009}
            \item Elephant56~\citep{salzaPasqualesalzaElephant562021}
            \item Spark-GA~\citep{maqboolScalableDistributedGenetic2019}
        \end{itemize} &
        Genetic operations are fitted to data-flow of big data paradigms, achieving high throughput and scalability &
        \begin{itemize}[left=0pt,label={\tiny\textbullet},itemsep=0pt,topsep=0pt,partopsep=0pt,nosep]
            \item Rigid communication patterns
            \item Limited support for external tools and workflows
        \end{itemize} \\
        \addlinespace
        Microservice &
        \begin{itemize}[left=0pt,label={\tiny\textbullet},itemsep=0pt,topsep=0pt,partopsep=0pt,nosep]
            \item AMQPGA~\citep{salzaSpeedGeneticAlgorithms2019a}
            \item KafkEO~\citep{mereloguervosIntroducingEventBasedArchitecture2018}
            \item BeeNestOpt.IAI~\citep{khalloofGenericFlexibleScalable2021}
        \end{itemize} &
        Container virtualization and stateless functions enable quick scaling and flexible software stacks in distributed cloud environments &
        \begin{itemize}[left=0pt,label={\tiny\textbullet},itemsep=0pt,topsep=0pt,partopsep=0pt,nosep]
            \item Proof of concept
            \item Not open source
            \item Incompatible with SLURM
        \end{itemize} \\
        \bottomrule
    \end{tabularx}
\end{table*}

\section{Related Works}\label{sec:PS}
Parallelism in evolutionary algorithms has been a research area closely coupled to the advances in compute hardware~\citep{albaParallelismEvolutionaryAlgorithms2002}.
A variety of parallel genetic algorithm frameworks have been developed in recent years. To provide a structured overview, we introduce a novel taxonomy that classifies existing frameworks based on their architectural design into three categories: monolithic, functional/data-flow-based, and microservice-oriented approaches. As summarized in Table~\ref{tab:framework_comparison}, this classification captures key differences in modularity, parallelism, and deployment strategy.

Monolithic frameworks, such as DEAP~\citep{derainvilleDEAPPythonFramework2012}, Geneva~\citep{berlichGenevaManual2025}, and Pagmo~\citep{biscaniParallelGlobalMultiobjective2020}, tightly integrate all components (e.g., initialization, propagation, fitness evaluation) within a single program, offering efficient resource utilization. Parallelism is often achieved through low-level implementations such as multiprocessing, multi-threading, message passing interface (MPI) based communication or shared-memory message passing~\citep{pandaMVAPICHProjectTransforming2021,wessnerParametricOptimizationHPC2023}.
These tightly coupled approaches excel at performance but are difficult to generalize and are typically limited with respect to workflow flexibility, utilizing distributed and heterogeneous hardware and the use of external tools.  

Functional and data-flow based approaches, including FlexGP~\citep{sherryFlexGPGeneticProgramming2012}, Offspring~\citep{vecchiolaMultiObjectiveProblemSolving2009}, Elephant56~\citep{salzaPasqualesalzaElephant562021}, and Spark-GA~\citep{maqboolScalableDistributedGenetic2019}, map genetic operations to established big data paradigms. Patterns such as MapReduce enable high throughput and robustness by acting on data through higher level or mathematical function without maintaining an inner state, treating data as immutable.~\citep{saltoBigOptimizationGenetic2023} While throughput and dataflow-centric design principles offer clear advantages, they introduce distinct drawbacks: they require specific data structures, and mapping complex algorithmic patterns to the frameworks native functionality is non-trivial.
For instance, MapReduce is a suitable scheme for mapping each individual to a worker and reducing all results into a common list. However, to handle dynamic mutation or crossover rates, complex analysis over the whole population must be performed to determine the next steps, thus breaking the predefined data flow and drastically reducing efficiency and throughput. Furthermore, the given functionality often relies on API implementations being available, limiting the compatibility across languages.

\begin{figure*}[ht]
    \centering
    \includegraphics[width=0.9\linewidth]{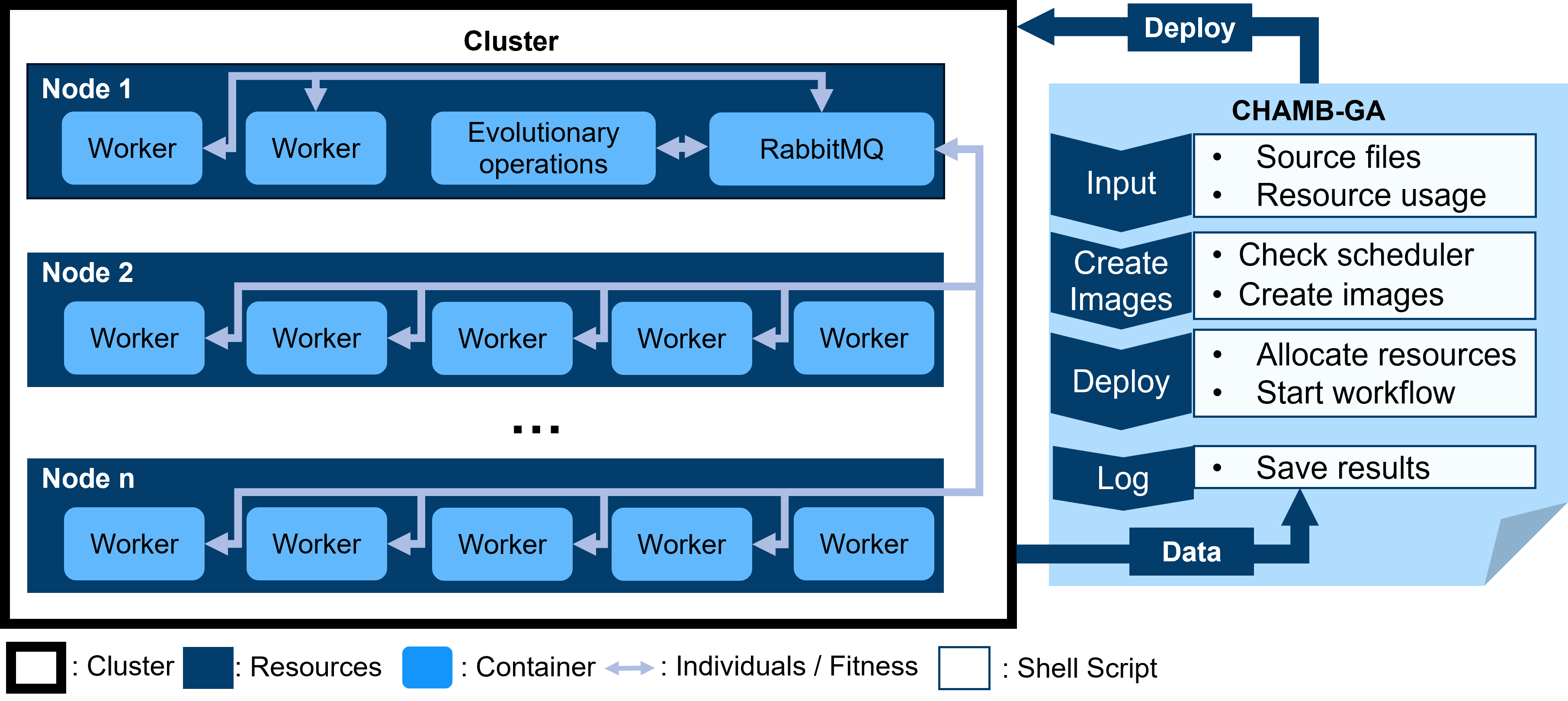}
    \caption{Overview of CHAMB-GA scripts (right), the cluster (left), and the involved data flows for an evolutionary optimization with distributed fitness evaluations. CHAMB-GA handles the user input, image build process and deployment, and manages the workflow, while the cluster handles the computational load without direct user interaction. }
    \label{fig:CHAMB-workflow}
\end{figure*} 

Microservice-based frameworks, such as AMQPGA~\citep{salzaSpeedGeneticAlgorithms2019a}, KafkEO~\citep{mereloguervosIntroducingEventBasedArchitecture2018}, and BeeNestOpt.IAI~\citep{khalloofGenericFlexibleScalable2021}, leverage containerization and stateless architectures to achieve modularity with communication through lightweight network technologies. Parallelism is handled by splitting the workload into separate tasks that can be scaled independently. The microservice approach is widely adopted in the area of cloud computing and is designed from ground-up as a distributed system achieving extensive scalability.
However, existing frameworks lack native support for the batch scheduling mechanisms of traditional HPC environments~\citep{10.1007/10968987_3, mcluckieGoogleCloudPlatform2014}, limiting their deployment on scientific computing clusters~\citep{ahnFluxOvercomingScheduling2020}.

While a wide variety of parallel GA frameworks exist, CHAMB-GA extends these approaches with its seamless integration of container orchestration and SLURM managed HPC clusters. This interoperability is critical as it bridges the gap between cloud-native flexibility and the computational power of traditional research clusters, while maintaining modularity through microservices. This seamless blend of technologies allows for integration of existing workflows extending the rigid scaling of monolithic approaches and the fixed patterns of functional and data-flow based approaches. CHAMB-GA's compatibility with SLURM managed HPC clusters allows for deployment on a wider variety of computational hardware. 
We demonstrate CHAMB-GA's massive scalability by scaling up to 25 nodes (3200 CPU cores) to optimize challenging AC powerflow problems, thereby providing a more robust empirical evaluation compared to proof-of-concept studies commonly found in current open-source microservice GA literature, see, e.g., ~\citet{mereloguervosIntroducingEventBasedArchitecture2018,salzaSpeedGeneticAlgorithms2019a}.

\section{Framework design}

\begin{figure*}[ht]
    \centering
    \includegraphics[width=0.9\linewidth]{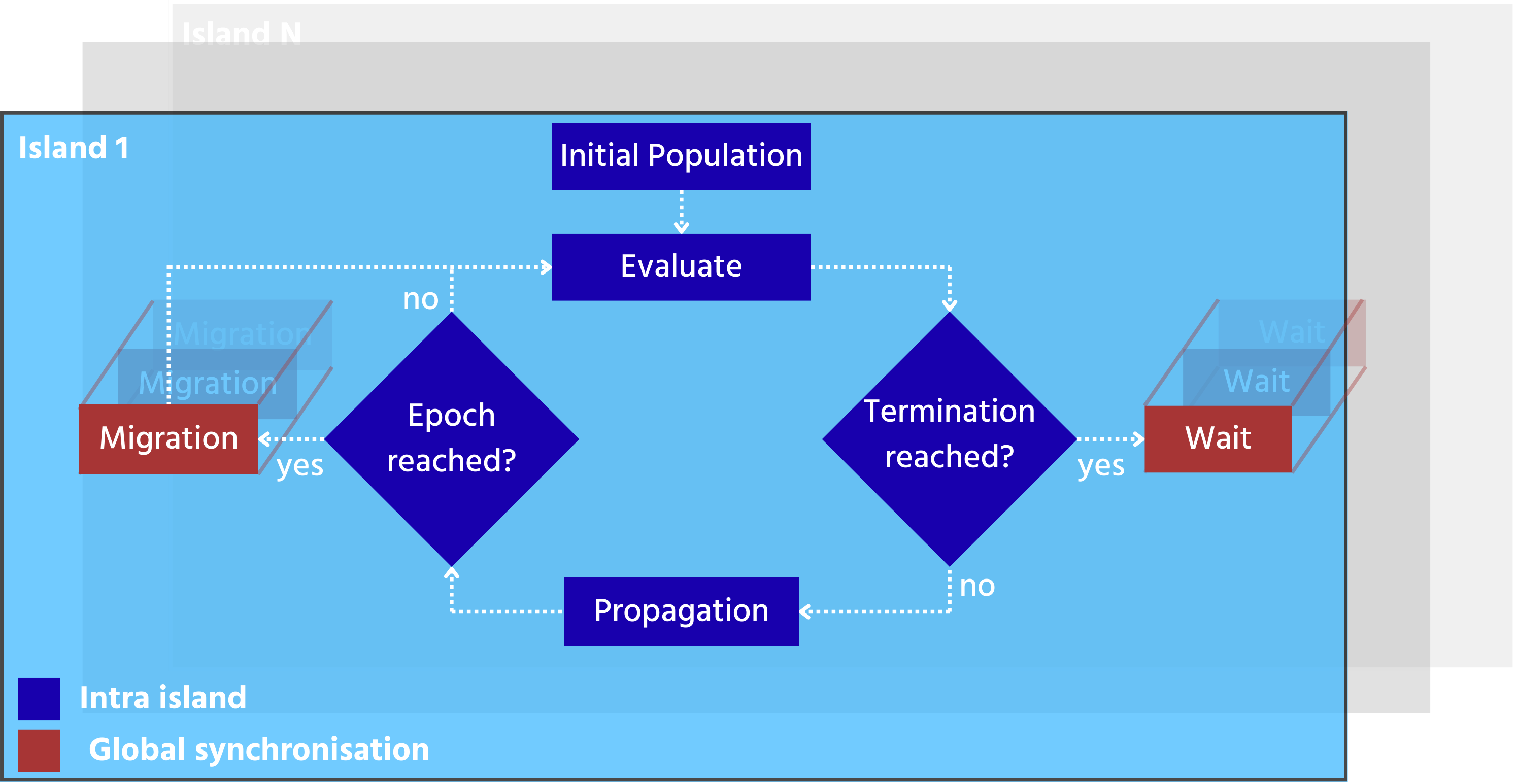}
    \caption{High-level overview of the evolutionary process in CHAMB-GA: Several islands operate independently, interacting only for migration and for global termination. Independent intra island execution steps are shown in blue; global tasks requiring synchronization (shown in red).}
    \label{fig:GA-workflow}
\end{figure*} 

CHAMB-GA is an orchestration script for massively scaling evolutionary algorithms. By integrating containerization with asynchronous communication, the framework physically decouples simulation workloads from the evolutionary propagation within a microservice architecture. Through seamless interaction with dynamic cloud-native orchestration (Kubernetes) as well as traditional HPC batch schedulers (SLURM), it ensures massive parallelism and reproducibility of optimization workflows across diverse distributed hardware environments. 

CHAMB-GA's structure is visualized in Figure~\ref{fig:CHAMB-workflow} for deployment of an optimization with distributed fitness evaluations on multiple CPUs across distinct nodes within a cluster.

The CHAMB-GA scripts act as a control hub that handles all user interaction. They automatically detect the underlying computational environment (Kubernetes or SLURM), facilitate image generation, orchestrate container deployment, and centralize both logging and data retrieval. 
Users interact with CHAMB-GA exclusively through a configuration file in JSON format. This file specifies the paths to the evolutionary algorithm and simulation source files, alongside the required compute resources, memory resources, and the desired scaling. CHAMB-GA handles source files consisting of a requirements file, a supplemental data folder, and a primary execution file. 
Utilizing these inputs, language specific, parameterized Docker files generate the required Apptainer ~\citep{kurtzerHpcngSingularitySingularity2021} (in case of SLURM) or Docker~\citep{merkelDockerLightweightLinux2014} (in case of Kubernetes) images tailored to the detected cluster architecture. 

By packaging the evolutionary operations and the simulation as independent containers, CHAMB-GA enables efficient and reproducible execution of complex distributed simulations~\citep{zhangSingularitybasedContainerTechnology2017}.
Once, the images are built, it orchestrates the deployment to the cluster. A strict sequence is enforced, first setting up the environment, then initializing the RabbitMQ message broker~\citep{RabbitMQOneBroker2025}, and finally the user-defined containers. To account for the separate file system within a container, CHAMB-GA handles the logging and retrieval of the final output from the cluster.

The cluster handles the computational load and executes the optimization workflow.
The basis of this workflow is a microservice architecture connecting all components via a central RabbitMQ message broker~\citep{RabbitMQOneBroker2025}.
Communication is unified through queues, accessible for all components in the cluster, allowing for the asynchronous distribution of individuals across all available simulation workers. After deployment, each container is assigned its specified resources and acts independently, communicating only with the message broker. This architecture is well suited for distributed island model GAs, as a shared queue acts as load-balance mechanism between the independent sub populations~\citep{izzoGeneralizedIslandModel2012}. Unlike traditional GAs that rely on rigid synchronization after each generation~\citep{biscaniParallelGlobalMultiobjective2020}, CHAMB-GA allows islands to operate independently in parallel, interacting with each other only for migration at the end of an epoch and global termination. This removal of synchronization barriers, enabled by asynchronous message passing, significantly increases parallel efficiency, especially if heterogeneous evaluations times would otherwise stall the evolutinoary process.
The general workflow of a distributed evolutionary process using multiple islands is shown in Figure~\ref{fig:GA-workflow}. 

With each island distributing its individuals to a shared set of workers, the evolutionary process of one island can be performed in parallel with the fitness evaluation of other islands.
CHAMB-GA mitigates the impact of heterogeneous fitness evaluation durations by evaluating queued individuals from a collective pool maintaining high resource utilization regardless of individual simulation time variance and thus maximizing throughput.

The architecture of CHAMB-GA accommodates complex, multi-level workflows, such as hyperparameter tuning or bilevel optimization, where an outer genetic algorithm dynamically communicates with an inner optimization loop or multi-stage simulation pipeline. This nesting is technically demanding as all components must execute in isolation and on demand, while the framework routes messages bidirectionally across hierarchical levels. To prevent resource contention, this isolation includes the source files, the software stack and the allocated compute resources, while the communication must avoid deadlocks during execution.

\begin{figure}[ht]
    \centering
    \includegraphics[width=0.99\linewidth]{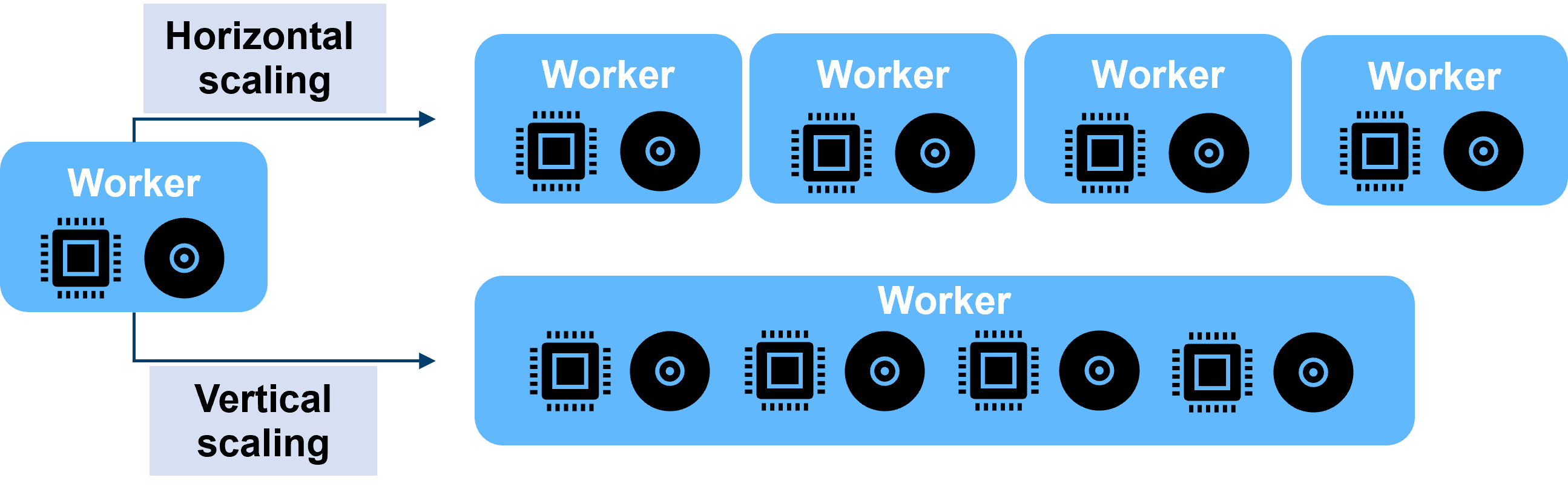}
    \caption{Horizontal and vertical scaling: Each block indicates a container and the assigned resources for CPU and memory are indicated by the amount of pictograms in each container.}
    \label{fig:vertical_horizontal_scaling}
\end{figure}

To allow for granular resource assignment within these nested structures, CHAMB-GA supports vertical and horizontal scaling (see Figure~\ref{fig:vertical_horizontal_scaling}). 
Horizontal scaling increases the throughput of individuals by increasing the amount of workers being deployed in parallel. Vertical scaling may decrease the time taken per fitness evaluation by increasing the resources available to each worker.

\begin{table*}[ht]
    \centering
    \caption{Computational hardware used for numerical results, along with the number of nodes utilized, the cores per node and the scheduling environment inherent to the system.}
    \label{tab: hardware tiers}
    \begin{tabular}{cccc}
        \toprule
        Tier  & \#Nodes & \#Cores per node & Scheduling environment\\
        \midrule
        single-node Kubernetes & 1 & 18 & Container orchestration (Kubernetes) \\
        multi-node Kubernetes & 3 & 128 & Container orchestration (Kubernetes) \\
        JURECA-DC & 28 & 128 & Batch scheduler (SLURM) \\
        \bottomrule
    \end{tabular}
\end{table*}

The optimal scaling strategy depends on the relationship between population size and the computational demand of a fitness evaluation as well as the available resources. Optimization problems requiring large populations align well with horizontal scaling, especially if the computational demand of the fitness evaluation is limited. If small populations are advantageous, the available compute resources can be used to vertically scale each worker. This dual scaling capability enables CHAMB-GA to better utilize available hardware and increase parallel efficiency for a wide variety of optimization problems.
\section{Numerical experiments} \label{sec:numerical_experiments}
We demonstrate the capabilities of CHAMB-GA in three steps.  The first step benchmarks framework scalability and overhead without computational load. The second step optimizes the dispatch of high voltage direct current (HVDC) lines under security constraints. The third step combines horizontal and vertical scaling with a hierarchical analysis consisting of an inner and an outer optimization loop for hyperparameter tuning. Steps 2 and 3 are computationally intractable without large-scale parallelism.

To guide the parallel evaluation, a slightly modified Non-Dominated Sorting Algorithm (NSGA-2)~\citep{debFastElitistMultiobjective2002} with islands and single-objective sorting is used. The total population size is distributed across $I$ islands with $P$ individuals each. The islands communicate via a ring migration pattern after $M$ generations (epoch) by sending out the best individual and replacing a randomly selected individual.

The ability to efficiently execute global optimization workflows across differently-sized compute hardware is demonstrated by scaling evaluations on a local computer, a multi-node Kubernetes cluster, and the SLURM managed JURECA-DC supercomputer of Forschungszentrum Jülich using up to 3500 cores across 28 nodes. The three considered hardware tiers are summarized in Table~\ref{tab: hardware tiers}. This tiered hardware selection demonstrates the portability and scheduler-agnostic deployment of CHAMB-GA on all sizes of hardware managed by either SLURM or Kubernetes.
In CHAMB-GA, independent processes are mapped to physical compute resources through the respective scheduler. Communication across distinct processes on the same node is fundamentally different compared to inter-nodal communication, making it complex to scale simultaneously across both. The queue based communication enables robust and flexible communication on all tiers, essential for analysis transitioning from local development to massively parallel evaluations. 

The lowest tier of computational power and complexity is a single-node Kubernetes cluster initialized with 18 CPU cores of an AMD EPYC 9534 64-core processor and 16 GB of RAM. The next tier is a dedicated multi-node Kubernetes cluster consisting of 3 nodes housing 128 cores split between two AMD EPYC 7763 processors and 256 GB RAM each. Rocky Linux 9.6 is used as host operating system while K3S v1.29.6+k3s2 is the installed Kubernetes version and 1.7.17-k3s1 the container runtime. The highest tier being the JURECA-DC supercomputer at the Forschungszentrum Jülich, consisting of 480 standard compute nodes each containing two 64 core AMD EPYC 7742 processors and 516 GB of RAM connected through NVIDIA’s InfiniBand HDR100~\citep{JURECAUserDocumentation}. CHAMB-GA is used to perform non-synthetic evaluations across up to 25 nodes, accumulating to 3200 CPU cores. This vastly exceeds the 4 nodes, 128 cores maximum of~\citep{khalloofGenericFlexibleScalable2023} and the 400 cloud based CPU cores of~\citep{sherryFlexGPGeneticProgramming2012}. 

\subsection{Baseline efficiency}\label{subsec: baseline efficiency}

\begin{figure*}[t]
    \centering
    \begin{subfigure}[b]{0.32\textwidth}
        \includegraphics[width=\linewidth]{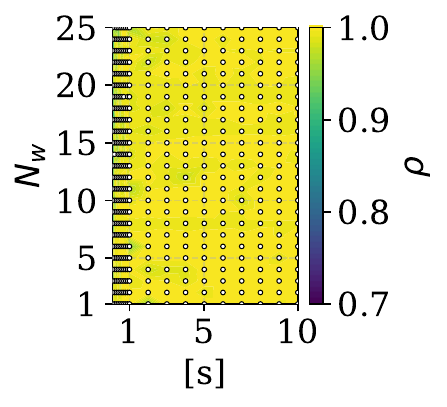}
        \caption{Single-node Kubernetes cluster (1-25 workers, 0.1-10 s evaluations).\\}
        \label{fig:Minikube_scaling}
    \end{subfigure}\hfill
    \begin{subfigure}[b]{0.32\textwidth}
        \includegraphics[width=\linewidth]{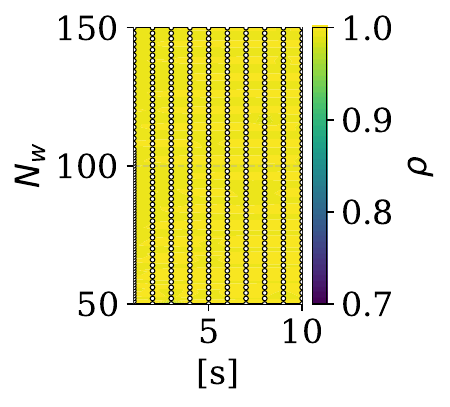}
        \caption{Multi-node Kubernetes cluster (50-150 workers, 1-10 s evaluations).\\}
        \label{fig:BORG_scaling}
    \end{subfigure}\hfill
    \begin{subfigure}[b]{0.32\textwidth}
        \includegraphics[width=\linewidth]{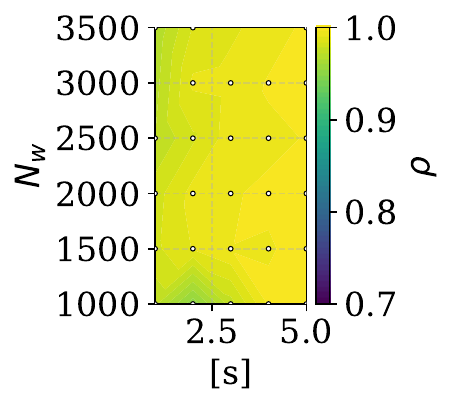}
        \caption{JURECA-DC supercomputer (1000-3500 workers, 1-5 s evaluations).}
        \label{fig:Jureca_scaling}
    \end{subfigure}
    \caption{Scalability tests to evaluate the parallel efficiency across three distinct computing environments: (a) a single-node Kubernetes cluster, (b) a multi-node Kubernetes cluster, and (c) the JURECA-DC supercomputer at Forschungszentrum Jülich for varying worker counts and fitness evaluation durations.}
    \label{fig:combined_scaling}
\end{figure*}

To assess the scaling efficiency of the CHAMB-GA framework, we conduct a baseline analysis across the three distinct hardware tiers. Isolating framework related overhead from variations in the fitness evaluation, we simulate the computational load using a sleep function of duration s, following~\citet{wessnerParametricOptimizationHPC2023}.
The parallel efficiency ($\rho$) is calculated by dividing the theoretical runtime of the GA by the measured wall-clock time (T):
\begin{align} \label{eq: parallel efficiency}
    \text{$\rho$} &= \frac{\text{s} \cdot \text{P} \cdot \text {M} \cdot \text{N}_E \cdot \text{I}}{\text{T}\cdot \text{N}_w}
\end{align}
where $\text{N}_E$ is the number of epochs and $\text{N}_w$ the number of parallel workers. As no computational load is considered, each worker is assigned a single CPU core.
The population size is adjusted alongside the number of parallel worker containers to guarantee a total computational load of at least 100 individuals per worker. From the theoretical runtime considered in Equation \ref{eq: parallel efficiency} follows that the population size is $P=\frac{\text{N}_w \cdot 100}{\text{I} \cdot \text{M} \cdot \text{N}_E}$, with P rounded up to the next integer. This ensures that the time spent on fitness evaluations significantly outweighs the one-time costs of deployment to the cluster and worker start-up, focusing on the communication overhead incurred by the framework. 
On each hardware tier a distinct set of combination of worker counts and sleep times is evaluated to reflect the scale of each tier. The worker counts range from a single worker on the single-node Kubernetes configuration up to $3500$ parallel workers on the JURECA-DC supercomputer, while the sleep time is varied from $0.1$ seconds up to $10.0$ seconds. 

The resulting parallel efficiencies for all combinations, averaged over 5 runs, are shown in Figure~\ref{fig:combined_scaling}. 
The lowest efficiency of $0.8$ is only recorded for the shortest evaluation times of $0.1$ seconds and $18$ parallel workers, shown in the top left of Figure~\ref{fig:Minikube_scaling}. The efficiency is above $0.95$ for all sleep times above $0.1$ seconds with up to 150 workers using Kubernetes as orchestrator (Figure~\ref{fig:BORG_scaling}) and up to 3500 parallel workers using SLURM as scheduler (Figure~\ref{fig:Jureca_scaling}). 

The deployment of the workflow onto the JURECA-DC compute cluster yields performance characteristics consistent with the Kubernetes-based results. Deploying CHAMB-GA utilizing SLURM and Apptainer~\citep{kurtzerHpcngSingularitySingularity2021} introduces negligible overhead, even at the scale of several thousand workers compared to the Kubernetes based results.

These results validate the framework's ability to scale efficiently across the full hardware hierarchy: from multiple cores on a single CPU, to multiple CPUs within a node, and finally across distributed nodes in a large-scale cluster independent of Kubernetes or SLURM being used for deployment.

\subsection{HVDC Dispatch} \label{subsec: HVDC dispatch}

\begin{table*}[ht]
    \centering
    \caption{NSGA-2 hyperparameter settings for the two scaling approaches: Population size (P), mutation method, mutation probability ($\mu_{mut}$), mutation distribution index ($\eta_{mut}$), crossover method, crossover probability ($\mu_{cx}$), crossover distribution index ($\eta_{cx}$), migration pattern and migration frequency (M).}
    \label{tab:GA_settings}
    \begin{tabular}{ccccccccccc}
        \toprule
        \multirow{2}{*}{Scaling} & \multirow{2}{*}{P} & \multicolumn{3}{c}{Mutation} & \multicolumn{3}{c}{Crossover} & \multicolumn{2}{c}{Migration} \\
        \cmidrule(lr){3-5} \cmidrule(lr){6-8} \cmidrule(lr){9-10}
         &  & Method & $\mu_{mut}$ & $\eta_{mut}$ & Method & $\mu_{cx}$ & $\eta_{cx}$ & Pattern & M \\
        \midrule
        (a)  & 412 & Polynomial\textsuperscript{a} & 0.7 & 34.6 & SBX\textsuperscript{b} & 1.0 & 97.5 & Ring & 5 \\
        (b) & 16  & Polynomial\textsuperscript{a} & 0.5 & 90.2 & SBX\textsuperscript{b} & 1.0 & 5.2  & Ring & 6 \\
        \bottomrule
    \end{tabular}
    \begin{tablenotes}
      \small
      \item[a] Polynomial mutation as implemented in original NSGA-II algorithm~\citep{debFastElitistMultiobjective2002}.
      \item[b] Simulated Binary Bounded Crossover - a real-coded crossover operator with similar search power as the single-point crossover in binary-coded GAs~\citep{debFastElitistMultiobjective2002}.
    \end{tablenotes}
\end{table*}

To evaluate CHAMB-GA under computational load, nonlinear AC powerflow calculations are used as fitness evaluation~\citep{molzahnSurveyDistributedOptimization2017}. Building on our previous work~\citep{Pesch:780996}, a realistic representation of the German high-voltage transmission network, based on the grid topology defined in the 2012 network development plan~\citep{Feix_Obermann_Hermann_Zelter_2012}, is considered. The network consisting of $2715$ buses and $871$ generators connected by $5351$ lines, with the three-phase network simplified into a single phase substitute system is extended to include the planned $18$ HVDC lines~\citep{Pesch:780996}. These DC lines enable bidirectional point-to-point low loss bulk power transport within a continuous range of  either $[-1300,1300] $ MW or $[-2000,2000]$ MW~\citep{Feix_Obermann_Hermann_Zelter_2012}. Each powerflow can be controlled individually and is a degree of freedom within the dispatch decision.

The objective is to minimize total grid usage fees defined through the total power transmitted across all lines, i.e.,
\begin{equation}\label{eq: min grid usage}  
	\min_{\mathbf{x}} F(\mathbf{x}) = \sum_{i\in \mathcal{L}} P_{i}(\mathbf{x}),
\end{equation}
where $\mathbf{x}$ is the vector of the $18$ HVDC powerflow settings (optimization variables), $\mathcal{L}$ is the set of \emph{all} lines in the network, and $P_{i}$ represents the positive powerflow transported through line $i$. 

Based on this minimization problem, CHAMB-GA's core capabilities are demonstrated in two distinct case studies. The first highlights the combination of vertical and horizontal scaling by extending the fitness evaluation to consider security constraints, and the second deploys a hierarchical workflow tuning the GA hyperparameters via a governing GA.

An extensive GA-based optimization minimizing~\ref{eq: min grid usage} was conducted to define the best found solution of the HVDC dispatch problem. The gap to this fitness value is used as measure for the solution quality in the subsequent evaluations. A negative gap might be achievable as the reference has no optimality guarantees.

\subsubsection{Horizontal and vertical scalability} \label{subsubsec: horizontal scaling}
In real-world power system operations, any viable HVDC dispatch configuration must satisfy $N-1$ security, ensuring the grid remains stable even under a single line failure. For the given network, instead of a single AC powerflow calculation each fitness evaluation is extended to include $2004$ contingency cases (failure of any 220kV or 380kV line), increasing the computational load by $3$ orders of magnitude. As all contingency cases are independent of each other, CHAMB-GA's flexible resource allocation allows the workers not only to be scaled horizontally, but also vertically by increasing the utilized CPU cores of each worker.

This two axis scaling can be applied in a wide variety of combinations, enabling tailored resource utilization. Two possible combinations are further considered highlighting the relevance on evolutionary optimization: (a) prioritizing horizontal scaling with $384$ parallel workers with $8$ CPU cores each, and (b) prioritizing vertical scaling with $24$ parallel workers, each utilizing 128 cores. Both configuration consist of a total of $3072$ cores to calculate the fitness values. These configurations are matched with suitable GA hyperparameters to decrease the time of cores being idle, shown in Table~\ref{tab:GA_settings}.

The fitness of each individual is adjusted to account for violations of the thermal limits during any contingency case. Any line failure resulting in a violation of a thermal line limit is called critical. Driving the GA population towards robust solutions, a penalty of $10\%$ is added to the objective for each critical line. To avoid even close to critical solutions, a smaller penalty of $1\%$ is added if any line loading reaches a loading between $95\%$ and $100\%$. The penalized objective function $F'(\mathbf{x})$, extending~\ref{eq: min grid usage}, is defined as:

\begin{equation}
\begin{split}
    \min_{\mathbf{x}} F'(\mathbf{x}) &= F(\mathbf{x}) \cdot \Biggl[ 1 + \sum_{c \in \mathcal{C}} \biggl( 0.10 \cdot \mathbb{I}_{10\%}(\mathbf{x}, c) \\
    &\quad + 0.01 \cdot \mathbb{I}_{1\%}(\mathbf{x}, c) \biggr) \Biggr],
\end{split}
\end{equation}
where
\begin{equation}
    \mathbb{I}_{1\%}(\mathbf{x}, c) = 
    \begin{cases} 
    1 & \begin{aligned}[t]
        &\text{if } \exists i \in \mathcal{L} \text{ s.t. } \\
        &P_{i,c}(\mathbf{x}) \ge 0.95 \cdot P_{i}^{\max} \\
        &\text{and } \mathbb{I}_{10\%}(\mathbf{x}, c) = 0
        \end{aligned} \\
    0 & \text{otherwise}
    \end{cases}
\end{equation}
and
\begin{equation}
    \mathbb{I}_{10\%}(\mathbf{x}, c) = 
    \begin{cases} 
    1 & \begin{aligned}[t]
        &\text{if } \exists i \in \mathcal{L} \text{ s.t. } \\
        &P_{i,c}(\mathbf{x}) > P_{i}^{\max}
        \end{aligned} \\
    0 & \text{otherwise}
    \end{cases}
\end{equation}

Here, $\mathcal{C}$ is the set of all $N-1$ contingency cases indexed by $c$, and $F'(\mathbf{x})$ is the penalty incurred objective function for the contingency analysis. The powerflow on line $i$ under contingency $c$ is denoted by $P_{i,c}(\mathbf{x})$, while $P_{i}^{\max}$ represents the thermal limit (maximal line loading) for line $i$. High- and overloaded lines are captured through the indicator function $\mathbb{I}_{1\%}$ and $\mathbb{I}_{10\%}$, respectively.

\begin{figure}[H]
    \centering
    \begin{subfigure}[b]{0.45\linewidth}
        \centering
        \includegraphics[width=\linewidth]{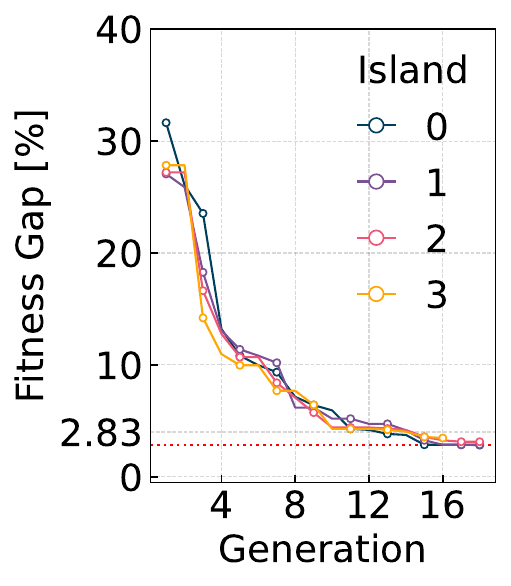}
        \caption{Best fitness value of each island in each generation for a GA evolving 384 individuals in parallel, each using 8 CPU cores with a time limit of 6 hours.}
        \label{fig:Contingency_GA_Gen_Hyper_parameters}
    \end{subfigure}
    \hfill
    \begin{subfigure}[b]{0.45\linewidth}
        \centering
        \includegraphics[width=\linewidth]{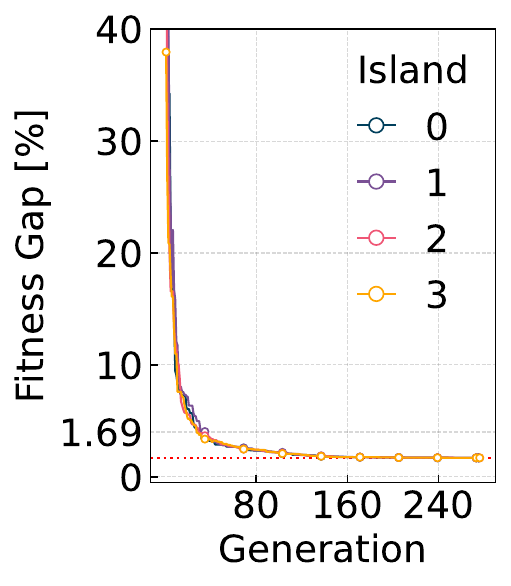}
        \caption{Best fitness value of each island in each generation for a GA evolving 24 individuals in parallel, each using 128 CPU cores with a time limit of 6 hours.}
        \label{fig:Contingency_GA_NFFE_Hyper_parameters}
    \end{subfigure}
    \caption{Best fitness of each island for the contingency constrained HVDC dispatch problem. The GA hyperparameters are set according to Table~\ref{tab:GA_settings} using a total of 3072 CPU cores for the fitness evaluation. The target fitness is a theoretical limit as it was achieved without penalization and without considering contingency cases.} 
    \label{fig:Contingency_GA_Solutions}
\end{figure}

Figure~\ref{fig:Contingency_GA_Solutions} shows the best found fitness of each island in each generation. Both configurations are evaluated for a wall-clock time of 6 hours, using the same amount of computational resources.
The best found solution in both configurations has no thermal line overloading, indicating a sufficiently large penalty.

During the evaluation of distributed GAs, scaling is often a trade-off between sampling efficiency and high throughput. While horizontal scaling is primarily limited by the GA itself through the population size, and the subsequent evolutionary pressure, vertical scaling efficiency often depends on the available parallelization of the fitness function. 
For the given security constrained powerflow based evaluation, a total of $60$ million AC powerflow calculations are evaluated for the primarily horizontal scaling in Figure~\ref{fig:Contingency_GA_Gen_Hyper_parameters}. For the vertically scaled approach in Figure~\ref{fig:Contingency_GA_NFFE_Hyper_parameters} the total number of evaluations decreased to roughly $36$ million.
While both number of evaluations are computationally intractable using a sequential or mildly parallel approach, this indicates that no single scaling approach is strictly superior.
Ultimately achieving a balance between sampling efficiency and optimization wall-clock time requires a flexible combination of both horizontal and vertical scaling to efficiently utilize available hardware. 

The results in Figure~\ref{fig:Contingency_GA_Solutions} show not only the capability, but also the necessity, for CHAMB-GA to allow for nuanced, problem and hardware specific adjustments for both vertical and horizontal scaling through user-input, without adjustments to the framework.

\noindent

\subsubsection{Hierarchical evaluation} \label{subsubsec: multi step}

CHAMB-GA enables flexible message routing and the deployment of multiple algorithm and simulation components to seamlessly interact with each other. A recurring hierarchical problem when using GAs is the choice of hyperparameters~\citep{maAutomatedAlgorithmDesign2026}, which, as shown in Section~\ref{subsubsec: horizontal scaling}, is problem specific and also depends on the available parallelism and target hardware.

To guide the hyperparameter search, we employ a hierarchical setup consisting of a governing GA (meta GA) and a set of worker GAs. Each individual of the meta GA encodes a hyperparameter configuration, whose fitness is evaluated by running a worker GA configured with those hyperparameters based on the AC powerflow HVDC dispatch problem. 
In principle, each worker GA could be parallelized internally (e.g., using Python's multiprocessing), but doing so would lock the degree of vertical parallelism at design time, an undesirable constraint given the large variations of the population size during the parameter search.

Instead, CHAMB-GA enables the deployment of a hierarchical workflow in which all three stages, (i) the meta-GA, (ii) the pool of worker GAs and (iii) the pool of AC powerflow evaluators, scale independently. The meta GA sends individuals (hyperparameters) to a pool of worker GAs, which in turn communicate with a pool of shared AC powerflow evaluators.

The shared pool design enables parallel evaluation while performing load balancing across worker GAs. Any idle evaluator picks up the next available individual, independent of its originating worker GA. This scheme scales more effectively than  statically assigning a fixed number of evaluators to each worker GA, especially when population sizes vary within the meta GA.

\begin{figure}[h]
    \centering
    \includegraphics[width=0.99\linewidth]{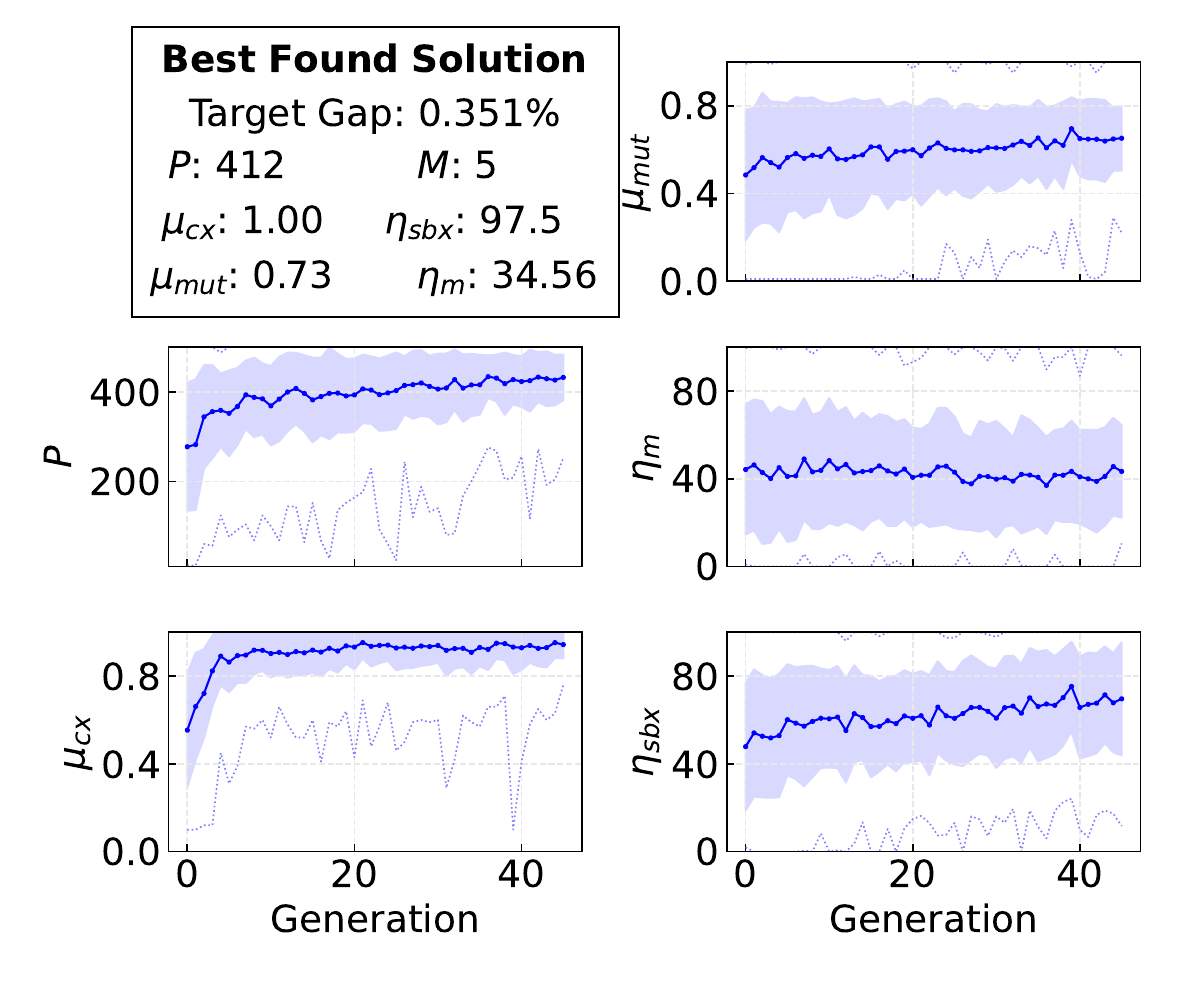}
    \caption{The best found parameters (top left) and the evolution of average hyperparameters within the population at each generation. The standard deviation of each parameter value within the meta GA generation is given as shaded area while the minimum and maximum values of the parameters are given as dotted lines. The meta GA is configured with $I=3$ islands and a population size of $P=96$. Each individual which corresponds to a set of hyperparameters is evaluated $5$ times, each time with a different seed for the worker GAs initial population.}
    \label{fig:hyperparameter tuning}
\end{figure} 

The genes of the meta GA are defined in Table~\ref{tab: Meta_GA_settings} together with their respective bounds. The genes of the worker GA are the 18 HVDC dispatch decisions without contingency considerations as defined in~\ref{eq: min grid usage}.
For the meta GA the termination criterion is a wall-clock limit of $12$~hours, while each set of hyperparameters is evaluated for 5 different seeds for 20 generations, and the overall best found solution is returned as fitness. 

\begin{table*}[ht]
    \caption{Definitions of the meta GA genes defining the hyperparameters tuned, the bounds define the search space. All parameters are implemented according to the NSGA-2 algorithm within the DEAP framework~\citep{derainvilleDEAPPythonFramework2012, debFastElitistMultiobjective2002}.}
    \centering
    \begin{tabularx}{\textwidth}{lcc>{\raggedright\arraybackslash}X}
        \hline
        \textbf{Hyperparameter} & \textbf{Symbol} & \textbf{Bounds} & \textbf{Description} \\
        \midrule
        Population size & $P$ & $[12, 500]$ & Number of individuals per island. \\
        Crossover probability & $\mu_{cx}$ & $[0.0, 1.0]$ & Probability of each individual to create an offspring by mating. \\
        Mutation probability & $\mu_{mut}$ & $[0.0, 1.0]$ & Probability of each individual to create an offspring through mutation. \\
        Mutation distribution index & $\eta_m$ & $[0.01, 100]$ & Crowding degree of the mutation operation, a high value indicating an offspring similar to the parents. \\
        Crossover distribution index & $\eta_{sbx}$ & $[0.01, 100]$ & Crowding degree of the crossover operation, a high value indicating an offspring similar to the parents. \\
        \bottomrule
    \end{tabularx}
    \label{tab: Meta_GA_settings}
\end{table*}

Figure~\ref{fig:hyperparameter tuning} shows the average value of each hyperparameter within a generation of the meta GA along with the standard deviation and minimal and maximal value within the same generation. 
These results provide insights into the evolutionary dynamics. $\mu_{cx}$ converges rapidly to its upper bound, whereas $\mu_{mut}$ increases slowly but exhibits a visibly shrinking standard deviation. Most notably, the population size remains well below its upper bound, indicating that, under the given termination criterion, horizontal scaling is inherently limited. Efficiently exploiting massively parallel compute resources therefore requires combining both scaling axes.

CHAMB-GA's native capability to combine horizontal and vertical scaling alongside multi-step workflows allows for optimization of a wide variety of problems. Complex workflows can be embedded into the optimization solely by adjusting the user input. Adding additional pools of workers can be used to, e.g., enable multi-oracle evaluations, to combine different simulation frameworks within a combined fitness evaluation or to allow for load-balanced vertical scaling even if it is not natively supported.

\section{Conclusion}

The growing demand for global optimization in scientific and engineering domains is progressively constrained by the computational cost associated with long-running simulations. Traditional optimization pipelines are tightly coupled to specific hardware or software stacks, limiting their scalability, applicability, and interoperability across distributed compute environments. CHAMB-GA addresses these issues using a containerized microservice approach that targets HPC scalability specifically. The proposed framework leverages a central message broker for asynchronous manager-worker communication, and seamlessly distributes evolutionary operations and fitness evaluations across heterogeneous infrastructure. It utilizes both SLURM's batch scheduling and Kubernetes container orchestration. This design eliminates the need to embed simulations within the same process as the genetic operations, enabling each component to be operated and scaled independently. 

The practicality of CHAMB-GA was demonstrated through a systematic benchmark study and a massively scaled case study that optimized HVDC line setpoints. The benchmarks confirm that, under both SLURM and Kubernetes, containerized orchestration with queue-based communication incurs negligible overhead. The system scales near-linear to more than $3500$ CPU cores distributed across $28$ compute nodes without reaching throughput saturation. These results validate the framework's ability to efficiently handle large-scale, asynchronous, evolutionary searches, independent of the underlying scheduler. This portability is essential for workflows transitioning from local development environments to massive production evaluations deployed on distributed systems and HPC clusters. 

When applied to a dispatch optimization problem derived from Germany's transmission grid extension plans, CHAMB-GA successfully transferred this scalability to an optimization under computational load. Including security relevant contingency cases increased the computational load by three orders of magnitude, evaluating nearly $60$ million AC powerflow calculations during a single optimization. This, sequentially intractable number of calculations required the framework's dual scaling capability. The GA's fitness evaluations scales horizontally while simultaneously each constrained powerflow calculation is scaled vertically. Together with CHAMB-GA's native ability to interact with SLURM- and Kubernetes-managed cluster environments, this enables problem-specific resource allocation tailored to the structure of each optimization task and the available hardware.

To further demonstrate the framework's flexibility and modularity, a hyperparameter analysis was conducted using a two-stage GA workflow. A meta GA that evolves worker GAs was configured to, in turn, interact with a shared pool of workers, without modifying the underlying architecture. Such hierarchical workflows, combining multiple algorithms and simulations with independent resource allocations, allow for complex optimization tasks to be incorporated. These tasks stretch from utilizing the queue-based load balancing across components to multi-oracle optimizations based on varying software and hardware stacks. 

CHAMB-GA advances the current state of distributed evolutionary algorithm frameworks by providing a unified, scalable, open-source approach. Its ability to adapt to diverse fitness evaluations or workflows as well as its seamless migration from Kubernetes to SLURM distinctively positions it as a widely applicable tool for massively scalable optimization. As a lightweight orchestration framework, it can be easily integrated with existing tools, allowing users to configure workflows through a single configuration file. 

Future development will focus on integrating heterogeneous computing resources, such as GPU acceleration, into the evolutionary pipeline. Additionally, the usage of heterogeneous programming languages needs to be evaluated to showcase the full flexibility and potential. 
\section*{Acknowledgment}
The authors gratefully acknowledge computing time on the supercomputer JURECA at Forschungszentrum Jülich under grant no. 60265 and financial support by the Helmholtz Association of German Research Centres through program-oriented funding.

\section*{Data availability}
The framework architecture is openly accessible via \href{https://jugit.fz-juelich.de/iek-10/public/optimization/chamb-ga#}{CHAMB-GA}. Results, algorithm implementation and simulation scripts are modified to exclude underlying power systems data, which are subject to confidentiality agreements and cannot be disclosed.

\section*{Declaration of competing interests}
The authors have no competing interests to declare that are relevant to the content of this article.

\section*{Authors' contributions}
Conceptualization: F.B., T.P., M.D.; Methodology: F.B, M.D., T.P.; Software: F.B.; Formal analysis and investigation: F.B.; Visualization: F.B.; Writing - original draft preparation: F.B.; Writing - review and editing: A.M., M.D., T.P., A.B.; Funding acquisition: A.M., A.B.; Supervision: A.M., M.D., T.P.

\section*{Declaration of generative AI and AI-assisted technologies in the manuscript preparation process.}
During the preparation of this work, F.B. used Gemini in order to correct grammar and spelling and to improve the style of writing. After using this tool, all authors reviewed and edited the content as needed and takes full responsibility for the content of the publication.
 
\bibliographystyle{elsarticle-num-names} 
    \renewcommand{\refname}{References}   
    \bibliography{bibexport}
\end{document}